\documentclass{emulateapj}
\newcommand{\abs}[1]{\left\vert#1\right\vert}
\newcommand{\be}{  \begin{eqnarray} }
\newcommand{\ee}{  \end{eqnarray} }

\newcommand{\beq}{\begin{equation}}
\newcommand{\eeq}{\end{equation}}

\newcommand{\bxi}{{\mbox{\boldmath $\xi$}}}

\newcommand{\grad}{{\mbox{\boldmath $\nabla$}}}

\def\simless{\mathbin{\lower 3pt\hbox
      {$\rlap{\raise 5pt\hbox{$\char'074$}}\mathchar"7218$}}}
\def\simgreat{\mathbin{\lower 3pt\hbox
      {$\rlap{\raise 5pt\hbox{$\char'076$}}\mathchar"7218$}}} %> or of order

\def\spose#1{\hbox to 0pt{#1\hss}}
\def\lta{\mathrel{\spose{\lower 3pt\hbox{$\mathchar"218$}}
     \raise 2.0pt\hbox{$\mathchar"13C$}}}
\def\gta{\mathrel{\spose{\lower 3pt\hbox{$\mathchar"218$}}
     \raise 2.0pt\hbox{$\mathchar"13E$}}}
\font\gkvec=cmmib10                         %for boldface lowercase Greek
\def\bxi{\hbox{{\gkvec\char24}}}           %bold face xi
\begin{document}
\title{Physical Basis for a constant lag time}
\author{Aristotle Socrates$^{*}$\& Boaz Katz$^{ *\dag}$ }
\affiliation{Institute for Advanced Study, Princeton, NJ 08540, USA}
\begin{abstract}
We show that the constant time lag prescription for tidal dissipation follows directly from 
the equations of motion of a tidally-forced viscous fluid body, given some basic assumptions.
They are (i) dissipation results from a viscous force that is proportional to the velocity of the 
tidal flow (ii) tidal forcing  and dissipation are  weak and non-resonant
(iii) the equilibrium structure of the 
forced body is spherically-symmetric.  The lag time is an intrinsic property of the tidally-forced body and is independent of the orbital configuration.  

\end{abstract}

\maketitle

\section{Introduction}

The origin of tidal dissipation in gaseous planets and stars is, for the most part, 
an un-solved problem.  Due to the extreme weakness of dissipation 
in these nearly perfect fluids, even identifying the correct theoretical
framework to address this issue has proven to be difficult.  Despite this, 
there has been a recent revival in tidal theory that has produced many 
promising results (Ogilvie \& Lin 2004;  Arras 2004; Wu 2005; Ivanov \& Papaloizou 2007;  
Goodman \& Lackner 2009;
Weinberg et al. 2012).    

In order to study the long-term evolution of orbits that are shaped by tidal 
dissipation, a parameterization of the strength of tidal dissipation is often performed.
One of the most common parameterizations is to incorporate a tidal lag time, where the forced response 
lags behind the equilibrium tidal deformation by some fixed value.  Hut (1981)
utilized this prescription in order to derive the orbit-averaged evolutionary equations
for a wide array of orbital configurations.  His results are widely used in many 
different astrophysical systems (e.g., Eggleton et al. 1998; Wu \& Goldreich 2002; 
Wu \& Murray 2003;
Fabrycky \& Tremaine 2007; Willems et al. 2010;
Wu \& Lithwick 2011; Socrates et al. 2012)

A benefit of utilizing the constant time lag parameterization is its ease of implementation.
That is, the constant time lag model may serve as a useful tool when inferring the strength of tidal 
dissipation of similar objects that reside in vastly different orbital configurations, which may
in constraining actual theories of tidal dissipation (see e.g., Socrates et al. 2012).  

In order to further this goal, we attempt to understand the underlying assumptions
behind the constant time lag model of Hut (1981).  In his analysis, Hut approximates that the induced 
quadrapolar tidal deformation as two point particles, where the line joining them spans
the diameter of the forced body and lags behind the line joining their center of mass and the distant source responsible for the tidal field.  Furthermore, the
mass of the point particles
is chosen such that the amplitude of their resulting quadrupole moment is equal to that of the equilibrium 
tide of a perfect fluid body.  A determination of 
the correspondence -- if any -- between the picture outlined above, 
to the fluid dynamics of a tidally forced body is the subject of this work.  

In \S\ref{s: basis} we demonstrate that, from the equations of 
motion, of a non-resonantly weakly forced, spherically symmetric 
body, the tidal response lags the equilibrium value by some fixed amount
of time, in the event 
that the dissipation is weak.  Furthermore, we show in \S\ref{s: basis}
that this constant lag time is an intrinsic property of the tidally forced body, in 
that it only depends upon its internal structure.  A brief discussion and summary are
given in \S\ref{s: summary}.

\section{Assumptions and Derivation of the Constant Lag Time model}
\label{s: basis}

\subsection{the tidal interaction}

Consider a spherical, self gravitating fluid object (star/planet) which is weakly perturbed by a slowly varying external gravitational field of the form
\begin{equation}
U_T(r,\theta,\phi,t)  = -\sum_{\ell m}r^{\ell}Y_{\ell m}(\theta,\phi)\,\Psi_{\ell m}(t).
\end{equation}
If, for example, the perturbation arises from a point mass $m_{\rm per}$ orbiting the object at distance $D(t)$ with angular coordinates $\theta'(t),\phi'(t)$ we have
\begin{equation}
\Psi_{\ell m}(t)= -Gm_{_{\rm per}}\sum_{\ell m}\frac{4\pi}{2\ell+1}\frac1{D(t)^{\ell+1}}
Y^*_{\ell m}\left(\theta '(t),\phi '(t)\right).
\end{equation}
To leading order in perturbation theory, the interaction potential is given by (cf. Newcomb 1962) 
\be
H_I=\int d^3x\,\rho\,\bxi\cdot{\grad} U_T
\ee
where $\int d^3x$ is taken over the volume of the forced body, $\rho({\bf x})$ is its unperturbed fluid density 
and $\bxi({\bf x},t)$ is the Lagrangian displacement field. 

We may write the interaction energy as
\be
H_I=-\sum_{\ell m}
q^*_{\ell m}\Psi_{\ell  m}
\label{e: H_I}
\ee
where $q_{\ell m}$ is the multipole moment (cf. Press \& Teukolsky 1977)
\be
q_{\ell m}\left(t\right)=\int d^3x\rho\,\bxi\cdot\grad r^{\ell}Y^*_{\ell m}\left(\theta,\phi \right).
\label{e: multipole}
\ee
Note that for $\ell=2$, the $\Psi_{\ell m}$'s have dimensions of frequency squared.

The rate of energy transfer $\dot{E}$ between the orbit and and the forced body is given by 

\be
\dot{E}=-\int d^3x\rho\,\dot{\bxi}\cdot\grad U_T = -\sum_{\ell m}\dot{q}^*_{\ell m}\Psi_{\ell m}.
\label{e: E_dot}
\ee
%The definition of $U_{T}$ in terms of the $\Psi_{\ell m}$'s may be employed in order to 
%write
%\be
%\dot{E}=-\sum_{\ell m}\dot{q}^*_{\ell m}\Psi_{\ell m}.
%\label{e: E_dot}
%\ee
Secular orbital evolution is entirely determined by the relationship between the the tidal 
potential, represented by the $\Psi_{\ell m}$'s, and the corresponding 
induced multipolar moments, $q_{\ell m}$'s.  That is, the gravitational potential induced by the perturbation of the body's mass distribution is fully determined by the $q_{\ell m}$'s. While the fluid response $\bxi$ may be complicated, only its multipolar moments affect secular orbital evolution.

\subsection{equation of motion, the equilibrium tide and higher-order corrections}

We neglect rotation when considering the equilibrium structure
as well as the dynamics of the fluid perturbations. Furthermore, we ignore the effects of rotation in the limit of non-resonant forcing.\footnote{Throughout, `non-resonant' refers to the fluid oscillations of the forced body.}  The tidal problem may be expressed as 
\be
\ddot{\mathbf{\bxi}} +{\bf C}\cdot\mathbf{\bxi}+{\bf D}\cdot\dot{\bxi}=-\grad U_T=-\sum_{\ell m}\Psi_{\ell m}(t)\,{\bf \Xi}_{\ell m}({\bf x})
\ee
where ${\bf C}$ is an Hermitian operator 
that is responsible for the restoring force (Lynden-Bell \& Ostriker 1967) and ${\bf D}$
is a time-independent differential operator that leads to dissipation.  In the co-ordinate
system of the forced body, the spatial dependance of the tidal forcing 
is given by the vector
\begin{equation}
{\bf \Xi}_{\ell\,m}({\bf x})=\grad r^{\ell}Y_{\ell m}\left(\theta,\phi\right),
\end{equation}
which for $\ell =2$ has dimensions of length.

If the time-dependance of the forcing is slow such that the inertia of the fluid is small, then
\be
\ddot{\mathbf{\bxi}}   \ll {\bf C}\cdot\bxi
\ee
and in the weak friction approximation
\be
{\bf D}\cdot\dot {\mathbf{\bxi}} \ll{\bf C}\cdot\bxi.
\ee
Given this ordering, we may approximate the solution as 
\be
\bxi\left({\bf x},t\right)=\bxi^{(0)}-{\bf C}^{-1}\cdot\left[{\bf D}\cdot\dot{\mathbf{\bxi}}^{(0)}
+\ddot{\mathbf{\bxi}}^{(0)}\right]
\label{e: bxi_expand}
\ee
and the equilibrium tide solution $\bxi^{(0)}$
is given by 
\be
\bxi^{(0)}\left({\bf x},t\right)=-\sum_{\ell m}\Psi_{\ell m}(t)\,{\bf C}^{-1}\cdot{\bf \Xi}_{\ell m}({\bf x}).
\label{e: bxi_sum}
\ee

\subsection{a single lag time for a given $\ell$}

The interaction energy, given by eq. \ref{e: H_I}, and the energy transfer rate, 
given by eq. \ref{e: E_dot}, only depend on the multipole moments $q_{\ell m}$.
%of the forced response.  In order to isolate the $q_{\ell m}$'s we project
%with ${\bf\Xi}_{\ell m}$ and integrate over the forced object
%\be
%\int d^3x\,\rho\,{\bf\Xi}^*_{\ell m}\cdot\bxi=q_{\ell m}(t).
%%-\tau_{\ell}\,\dot q^{(0)}_{\ell m}-P^2_{\ell}
%%\ddot q^{(0)}_{\ell m}.
%\ee
When computing the multipole moments $q_{\ell m}$'s by inserting eqs. \ref{e: bxi_expand}
and \ref{e: bxi_sum} into eq. \ref{e: multipole}, coefficients of the following form are
encountered
\be
{\rm M}_{\ell\ell' m m'}=\int d^3x\rho\,{\bf\Xi}^*_{\ell m}\cdot{\bf M}\cdot{\bf\Xi}_{\ell' m'}
\ee
where ${\bf M}={\bf C}^{-1}$ or ${\bf M}={\bf C}^{-1}\cdot{\bf D}\cdot{\bf C}^{-1}$.
By assuming that ${\bf C}$ and ${\bf D}$ are rotationally invariant, which is equivalent 
to assuming the equilibrium structure of the forced body is spherically symmetric, then 
\be
{\rm M}_{\ell\ell' m m'}={\rm M}_{\ell}\,\delta_{\ell \ell'}\delta_{mm'}.
\ee
It follows that we can express a given multipole moment to leading order as
\be
q_{\ell m}(t)=q_{\ell m}^{(0)}(t)-\tau_{\ell}\,\dot{q}_{\ell m}^{(0)}(t)-P^{-2}_{\ell}\,\ddot{q}_{\ell m}^{(0)}(t)
\ee
where a single lag time $\tau_{\ell}$ for a given $\ell$ is 
\be
\tau_{\ell}=\frac{\int d^3x\,\rho\, {\bf \Xi}^*_{\ell m}\cdot{\bf C^{-1}} \cdot{\bf D}\cdot{\bf C}^{-1}
\cdot{\bf \Xi}_{\ell m}}{\int d^3x\,\rho\,{\bf\Xi}^*_{\ell m}\cdot{\bf C}^{-1}
\cdot{\bf \Xi}_{\ell m}}
\label{e: lag_derive}
\ee
and $P^2_{\ell}$ is given by
\be
P^2_{\ell}=\frac{\int d^3x\,\rho\abs{{\bf C}^{-1}\cdot{\bf\Xi}_{\ell m}}^2}{
 \int d^3x\,\rho\,{\bf\Xi}^*_{\ell m}\cdot{\bf C}^{-1}
\cdot{\bf \Xi}_{\ell m}}.
\label{e: P_ell}
\ee
Though the integrals above contain the azimuthal quantum number $m$, spherical symmetry
requires invariance under rotations and consequently, independence of $m$.

The term $\propto \ddot{q}^{(0)}_{\ell m}$ does not lead energy or angular momentum transfer since
it contribution is a full derivative in time, which does not accumulate over long time-scales. The 
leading order terms of the multipole deformation $q_{\ell m}(t)$
that are responsible for apsidal precession and secular 
orbital evolution are thus given by
\begin{eqnarray}
q_{\ell m}\left(t\right)&=&q^{(0)}_{\ell m}\left(t\right)-\tau_{\ell}\,\dot q^{(0)}_{\ell m}\left(t\right)\simeq q^{(0)}_{\ell m}\left(t-\tau_{\ell}\right)
\end{eqnarray}
where we assumed that $\tau_{\ell}$ is small in comparison 
to the characteristic time in which the tidal potential varies.
Sole consideration of the quadrupolar $\ell=2$ response of the expression above is
equivalent to eqs. 2 \& 3 of Hut (1981), which together serve as the starting point and 
underlying assumption of his analysis.  Furthermore, by inserting the above expression 
for the $q_{\ell m}$ into eq. \ref{e: E_dot}, we may write the secular energy transfer rate 
as 
\be
\dot{E}=\sum_{\ell m}\,\tau_{\ell }\,\ddot{q}^{(0)*}_{\ell m}\Psi_{\ell m}.
\ee
Again, by restricting to $\ell=2$, the relation above is equivalent to eq. 40 of Eggleton et al. (1998),
which serves as the starting point of their derivation for the secular equations for orbital 
evolution.

The lag time $\tau_{\ell}$ is completely determined by the equilibrium 
structure of the forced body.  That is, $\tau_{\ell}$ is determined by $\rho({\bf x})$,
${\bf C}$ and ${\bf D}$ all of which,
under the assumptions previously mentioned, only depend upon the equilibrium 
structure of the object in question.  Similar conclusions can be deduced from the
analysis of Willem et al. (2010).  Those authors considered the case of a spherically
symmetric radiative star where dissipation results from thermal diffusion and turbulent
viscosity.

\subsection{constant density equilibrium structure}

As an illustrative example, consider a constant density equilibrium structure.  Reisenegger
(1994 and references therein) points out that for such an idealized system, ${\bf\Xi}_{\ell m}$
is an eigenfunction of ${\bf C}$, with eigenvalue $\omega_{\ell}$.  
Furthermore, and by construction, ${\bf\Xi}_{\ell m}$ is 
responsible for the entire multipolar response from the tidal acceleration $-\grad U_T\propto
{\bf\Xi}_{\ell m}$.  In this case, the equilibrium tide becomes
\be
\bxi^{(0)}_{\ell m}=-\frac{\Psi_{\ell m}(t)}{\omega^2_{\ell}}\,{\bf\Xi}_{\ell m}.
\ee
The displacement field ${\bf\Xi}_{\ell m}$ is fundamental mode of the forced 
body, which is sometimes referred to as the $f-$mode or the Kelvin mode.
${\bf\Xi}_{\ell m}$ has no radial nodes and its period of oscillation is, essentially, the 
free-fall time at the surface.

In this limit, the lag time $\tau_{\ell}$ becomes
\be
\tau_{\ell}=\frac{\gamma_{\ell}}{\omega^2_{\ell}}\equiv
\frac{\int d^3x\,\rho\, {\bf \Xi}^*_{\ell m}\cdot{\bf D}
\cdot{\bf \Xi}_{\ell m}}{\omega^2_{\ell}\,\int d^3x\,\rho\,{\bf\Xi}^*_{\ell m}
\cdot{\bf \Xi}_{\ell m}}
\label{e: tau_Reiss}
\ee
where $\gamma_{\ell}$ is the damping rate of the ${\bf\Xi}_{\ell m}$'s 
and
\be
P^2_{\ell}=\omega^{-2}_{\ell}.
\ee
Each ${\bf\Xi}_{\ell m}$ need not be an eigenvector of ${\bf D}$ in order for eq. \ref{e: tau_Reiss}
to be correct.  As in the more general case, the only requirement on ${\bf D}$ for 
producing a single constant time lag $\tau_{\ell}$ is for it to be invariant under rotations.
Finally, for a constant density equilibrium structure, the tidal problem is equivalent
to a set of decoupled forced damped harmonic oscillators with equal 
constant lag times.

\section{discussion and summary}
\label{s: summary}

The constant time lag model of Hut (1981) is perhaps the most widely 
used prescription for parameterizing the strength of tidal dissipation.
Hut states that the most attractive feature of the constant $\tau$ model for tidal dissipation is 
its simplicity.  Perhaps this is true.  However, as we have shown, the constant $\tau$
model for tidal dissipation follows from some very basic physical assumptions.  Namely, the 
tidal forcing non-resonant, the forced body is a spherically symmetric fluid and
the dissipation as well as the tidal forcing, is weak.
In fact, these assumptions are well-approximated in many astrophysical environments
and can be, in principle, tested with observations of stars (cf. Dong et al. 2012) 
and extra-solar planets.

We demonstrated that the lag time $\tau_{\ell}$ is an intrinsic property of the tidally-forced
object.  In other words, two identical objects placed in vastly different 
orbital configurations possess identical values for $\tau_{\ell}$ as long as the 
underlying assumptions of a constant $\tau_{\ell}$ remain valid.

For example, Socrates et al. (2012) show that in order for a Jupiter 
analogue to undergo high-$e$ migration, the required lag time must be at least ten times stronger
than that inferred from the Jupiter-Io interaction.  Therefore, either the 
assumptions that led to a constant lag time are either incorrect or, for example, another
element must be added to the theory of high$-e$ migration.

\acknowledgements{We thank Phil Arras,  Subo Dong, Peter Goldreich and Scott Tremaine for many 
helpful suggestions and conversations.  We also thank
John Papaloizou, Rafael Porto and Matias Zaldarriagga for stimulating discussions. B.K was
supported by NASA through Einstein Postdoctoral Fellowship awarded by
the Chandra X-ray Center, which is operated by the Smithsonian
Astrophysical Observatory for NASA under contract NAS8-03060. AS acknowledges
support from a John N. Bahcall Fellowship awarded by the Institute for Advanced
Study, Princeton.}

\end{document}